# Optimization of terahertz absorption in periodic quantum well structures

Zahra Javidi[1], Mehdi Hosseini[1a] and Mohammad Javad Karimi[1]

[1]*Department of physics, Shiraz University of Technology, Shiraz, 313-71555, Iran*

In this work, the linear absorption spectra for periodic arrays of GaAs-GaAsAl quantum wells with different thicknesses, inside electromagnetic wave has been studied. The eigen energies and eigen functions are calculated by solving the Schrödinger equation numerically. The absorption spectra are obtained using the density matrix approach and the effects of quantum well parameters have been studied. Results show that for a wide range of parameters the absorption peaks lie in the terahertz region. Furthermore, it is possible to adjust the frequency of absorption peak in the terahertz range by changing the width and height of the wells or array numbers that could be used in terahertz devices.

## I. INTRODUCTION

Today's, the semiconductor quantum structures including quantum wells, quantum wires, quantum dots, quantum emitters and super-lattices, due to their unique properties are the subject of many research [1-6]. The optical and electronic properties of these structures are of great interest due to their extensive functional capabilities in semiconductor optoelectronic devices, such as Infrared lasers, far-infrared photo-detectors, electro-optical modulators, and optical switches [7-9].In addition, due to the widespread use of semiconductor quantum wells (QWs) in the fabrication of electronic devices, optoelectronics, semiconductor lasers, infrared detectors, the study of the external factors on the optical properties of QWs systems is important [7-9].

---

[a] hosseini@sutech.ac.ir

So far, many studies have been carried out to find the linear and nonlinear absorption coefficients of QWs with different shapes [10-13]. The results of these works show that the absorption coefficients can be very sensitive to various parameters such as applied electric and magnetic fields, incident optical intensity, and the shapes of QWs [12, 14]. On the other hand, the terahertz absorber due to its wide applications and also lack of proper high-performance chips has been heavily investigated by scientists in recent years [15-18]. Usually, the frequency between 0.3 and 30 *THz* is called the terahertz range [18, 19]. Today's, terahertz technology is driven by scientific and industrial applications in areas such as information and communications technology [17], biology and medical science [21], security check and inspection, quality control in production and packaging, detectors, and so on [16-20]. Quantum wells and semiconductors are good candidates for absorber and sensors in the terahertz range [21-23].

In recent decades, multiple quantum wells have been investigated due to their potential applications (e.g., photo-detectors, high-speed absorption modulators, electro optical switches, and terahertz oscillators) [24]. Technological applications of multiple quantum wells depend on their optical and transportation properties that related to the quantum well specification [25]. The intersubband optical absorption in an asymmetric double quantum well for different barrier widths calculated by Ozturk et al. [26]. The results reveal that the intersubband transitions and the energy levels in an asymmetric double quantum well can be modified and controlled by the barrier and the well width. The sensitivity of the absorption coefficient to the barriers and wells width can be used in various optical semiconductor device applications.

Kasapoglu et al. investigated the effects of the geometrical parameters and electric field in a double inverse parabolic quantum well [24]. They indicated that by changing the *Al* concentrations in the center of the wells, the electric field and the geometric parameters of the well, absorption peak position could be tuning. Ungan et al. studied the linear and third-order nonlinear optical absorption coefficients in a modulation-doped asymmetric double quantum well [27]. Their results show that the total absorption coefficients shift toward higher energies as the right-well width decreases and the total optical absorption coefficients strongly depend on the incident wave intensity. Ozturk et al. examined the effect of the barrier widths on the linear and nonlinear optical absorption coefficient of triple quantum wells. Their results indicate the energy interval changes with barrier widths and the linear and nonlinear optical absorption coefficients depend on the barrier widths. Therefore, different optical absorption coefficients which may be suitable for various optical modulators and infra-red optical device applications are obtained by changing barrier widths [25].

In this work, we study the absorption spectrum of double quantum wells (DQW) and triple quantum wells (TQW) in the terahertz frequency range. The effects of the potential height of the barriers, wells width and the number of layers are investigated.

## II. THEORETICAL MODEL

As shown in Fig. 1, we consider a frequent array of GaAs/AlGaAs DQWs and TQWs which grown in the z-direction.

The corresponding confinement potential of these structures are as follows (Eq. (1 and 2)):

$$V_{DQW}(z) = \begin{cases} 0 & N(W_A + W_B) < Z < N(W_A + W_B) + W_A & \text{A Layers} \\ V_0 & N(W_A + W_B) + W_A < Z < (N+1)(W_A + W_B) & \text{B Layers,} \end{cases} \quad (1)$$

$$V_{TQW}(z) = \begin{cases} 0 & N(W_A+W_B+W_C) < Z < N(W_A+W_B+W_C)+W_A & \text{A Layers} \\ V_0 & N(W_A+W_B+W_C)+W_A < Z < N(W_A+W_B+W_C)+W_A+W_B & \text{B Layers} \\ V_0' & N(W_A+W_B+W_C)+W_A+W_B < Z < (N+1)(W_A+W_B+W_C) & \text{C Layers,} \end{cases}$$

(2)

where $N$ is the number of layers, $W_A$, $W_B$ and $W_C$ are the widths of the $A$, $B$ and $C$ layers, respectively. $V_0$ and $V_0'$ are the potential profile of the conduction-band which determines as follows (Eq. 3) [28]:

$$V_0 = 0.6\Delta E_g(x) = 0.6(1.155x + 0.37x^2)$$
$$V_0' = 0.6\Delta E_g(y) = 0.6(1.155y + 0.37y^2),$$

(3)

where $x$ and $y$ are A$l$ mole fraction.

In order to calculate the optical properties, the energy eigenvalues and Eigen functions are obtained by solving the Schrödinger equation numerically. The boundary conditions between layers are satisfied as well. By having the dipole moment matrix elements that is calculated from Eigen functions, the absorption spectra are obtained using the density matrix approach. Then, the effects of different parameters on the optical absorption of the system have been studied.

By considering the incident electromagnetic field as a time-dependent harmonic electric field as (Eq. 4):

$$\varepsilon(t) = \varepsilon_0 \cos\omega t = \tilde{\varepsilon} e^{i\omega t} + \tilde{\varepsilon} e^{-i\omega t},$$ (4)

in which $\omega$ is the frequency of the external incident field, the linear absorption coefficient according to the density matrix approach is given as follows (Eq. 5):

$$\alpha^{(1)}(\omega) = \sqrt{\frac{\mu}{\varepsilon_R}} \frac{|M_{21}|^2 \hbar \omega \Gamma_2}{(\varepsilon_{21}-\hbar\omega)^2 + (\hbar\Gamma_0)^2},$$ (5)

5where $\eta_v$ is the electron density, $\mu$ is the permeability, $E_{21}$ is the energy difference of electronic states, $M_{ij}$ is the dipole transition matrix element, and $\varepsilon_R$ is the real part of the permittivity [13].

## III. RESULTS AND DISCUSSION

The parameters used in calculations are as follows: $m = 0.067\, m_0$, where $m_0$ is the free electron mass, $\eta_v = 3.0 \times 10^{22}\, m\text{-}3$, $\gamma_{12} = 1/T_{12}$ where $T_{12} = 0.14\, ps$, [12].

### 3.1 A periodic array of double quantum wells

Here, the number of double layers is shown with $N$, so there are $2N$ layers for each value of $N$.

Fig. 2 shows the absorption spectrum as a function of frequency for some typical values of quantum well parameters. This figure reveals that the sharp absorption peak with the full width at half maximum less than one terahertz could be achieved that is suitable for device application. It is also clear that the absorption peak could be tuned over the wide range of terahertz frequency, by changing the physical parameters of the wells.

The geometrical parameters for Figs 3 and 4 are given in table 1.

Fig. 3 presents the maximum absorption coefficient (Fig.3-a) and position of frequency peak (Fig.3-b) versus the thickness of layers ($W_A=W_B$) and the number of layers. Fig. 3-a shows that the absorption peak decreases with increasing the thickness of B layers. In Fig. 3-b, the red color region corresponds to absorption peak more than 60 *THz*, whereas the blue color indicates absorption peak less than *20 THz*. This figure shows that by adjusting the number and thickness of layers, it is possible to tune the absorption frequency in the range of terahertz. Fig. 3-b also reveals that for $N<20$ and $W_B<20$, the frequency of absorption peak is out of terahertz range (>30*THz*) but by increasing the number of layers, the frequency peak could be decreases to less than 10 *THz*.

Fig. 4 shows the maximum of absorption coefficient and position of frequency peak versus the parameters of quantum wells. Fig. 4-a and 4-c illustrate that the absorption coefficient decreases slightly with increasing the thicknesses and for all thicknesses less than 30 nm, the absorption coefficient is more than $1.4\times10^4 m^{-1}$. The high absorption coefficient in all considered range of thicknesses and potentials is suitable and due to the robustness of the fabricated devices in the mentioned range of frequency. Fig. 4-b shows that for almost all of the considered ranges of parameters, the peak frequency is within the terahertz range. This figure reveals that by increasing the thickness, the absorption peak moves toward lower frequencies. Of course, this should be also noted that by increasing the thickness of the layers, the absorption decreases slightly, so the optimization needs for the desired application. This figure also indicates that the peak frequency is a function of the total thickness of the two barriers approximately and by increasing the total thickness of more than 30 nm, the peak frequency of 10 *THz* or smaller can be obtained. Fig. 4-c illustrate that by decreasing the thickness of the layers and the potential of the barriers, the peak of the absorption increases. Fig. 4-d reveals that by increasing the potential height of barrier *B*, the peak frequency increases. This figure also shows that the dependence of the peak absorption to the width of quantum wells is only noticeable in low thicknesses, and dependence almost disappears by increasing the thickness to 20nm. Although the dependence of the peak frequency to the barrier potential is negligible for the high thicknesses, it is recognizable at the low thicknesses. It should be noted that due to the dependence of the potential well to the concentration of aluminum, by changing the concentration of aluminum, the potential of the well can be changed.

**3.2 A periodic array of triple quantum wells**

The studied quantum well structure is consisting of the *GaAs*, *GaAs$_x$Al$_{1-x}$* and *GaAs$_y$Al$_{1-y}$*. The number of triple layers is shown with *N*, so that for specific *N* there are 3*N* layer. The physical parameters of the next figures are presented in the table 2.

Fig. 5 shows the peak values of the absorption coefficient and its frequency versus the potential height of barrier *B* and the number of layers. It is clearly seen from Fig. 5-a that the values of absorption are higher than $5\times10^4$ m$^{-1}$. This figure also indicates that the maximum of absorption is occurred for the 8 triple of layers and barrier height less than 0.2eV. It is also seen that by decreasing the potential height of the barrier, the maximum of absorption increases but by increasing the set of layers more than 10, the maximum of absorption decreases. Fig. 5-b reveals that the peak frequency within terahertz range (<30*THz*). This figure illustrates that by increasing the set of layers more than 10, the peak frequency of 10 terahertz or smaller can be obtained. This figures also shows that the frequency of absorption peak is almost independent on the $V_0$.

Fig. 6 shows the peak of the absorption coefficient and the frequency versus the thickness of layer *B* and the number of layers. In Fig. 6-a, it is clearly seen that the values of absorption for nearly all regions are higher than $5.4\times10^4$ m$^{-1}$. Fig. 6-b reveals that the peak frequency within terahertz range (<30*THz*) for the number of layers above 18 (*N*>6), also the peak frequency of about 5 *THz* could be obtained for *N*>10.

Fig. 7 shows the maximum values of the absorption coefficient and their frequencies versus the potential height of barriers *B* and *C*. These figures show that the absorption peak is higher than $5.2\times10^4$ m$^{-1}$and its variation is less than 10% therefore it is suitable for terahertz sensors application. It is seen that there is a point in each plot (that may not be observed in some of them for considered parameters) that the absorption peak frequency is maximum and by increasing the distance from this point the absorption peak frequency decreases. In Fig. 7-d this maximum is lied in the range of parameters and it is clearly observable at height of the barrier *B* and *C* about 0.18eV. By comparison of Figs. 7-a, 7-c and 7-e, it is clear that the absorption peak is almost independent of thickness and potential of the barriers. But the peak frequency is severely dependent to thickness of the layers and weakly dependent to the potential of the barriers. Due to high values of the absorption (higher than $5.2 \times 10^4 m^{-1}$) in all the considered range and the small variation of the peak frequency versus the potential of the barriers and major variation of the peak frequency versus thickness of the layers inside the terahertz range (<20*THz*), this figures reveal the promised tools for fabrication and tunability or fine adjustment of a terahertz sensor even for frequencies smaller than 10 *THz*.

Fig. 8 shows the maximum value of the absorption coefficient and its frequency versus the thickness and the potential height of the barrier *B* for the different layer thicknesses. Fig. 8 indicates that by decreasing the thickness of the layers and the potential of the barriers, the peak of the absorption increases and reaches to $5.7 \times 10^4 m^{-1}$. Furthermore, this figure reveals that by increasing the layers thickness the frequency peak shifts toward the smaller frequencies in the terahertz range. This figure illustrates that the thickness of the layers has a significant effect on the frequency peak but the barriers height has a negligible effect. However, by increasing the thickness, the value of absorption decreases, therefore an optimization is required for specified application.

Fig. 9 reveals the peak of the absorption coefficient and frequency peak versus the thickness of layers *B* and *C*. Fig. 9-a and 9-c illustrate that by decreasing the thickness of layers *B* and *C*, the absorption increases. By comparison between these two figures, it could be concluded that the absorption decreases by increasing the set of layers. Fig. 9-b and 9-d show that by increasing the thickness of layers *B* and *C*, the frequency decreases within the terahertz range, it is also clear from Fig. 9-d that the frequency peak around 4 *THz* could be achieved for 14 set of layers. These figures indicate that the frequency peak is a function of the total thickness of the both barriers.

The absorption peak value and its frequency versus the potential height of barriers *B* and *C* is depicted in Fig. 10. It can be seen from Fig. 10-a and 10-c that by decreasing the potential height of barriers *B* and *C*, the peak of absorption increases. Fig. 10-b and 10-d illustrate that by increasing the potential height of barrier *B* and by decreasing the potential height of barrier *C*, the frequency peak decreases till *8 THz*. By increasing the set of layers (in Fig. 10-a and 10-c) the absorption decreases slightly and (in Fig. 10-b and 10-d) the frequency peak decreases severely.

The figures for 24 and 42 layers show that the absorption is high and stable enough and the frequency peak is inside that terahertz range. Also by changing the potential of barriers *B* and *C* it is possible to fine tuning the frequency of absorption peak.

## IV. CONCLUSION

In this work, we have studied the terahertz absorption in double and triple periodic quantum wells. The obtained results show that the frequency of the absorption peak is within the terahertz range and it can be adjusted by changing the thickness, *Al* concentration and the number of layers. The results reveal that by increasing the thickness and the set of layers, the peak frequency decreases while by increasing the potential height, absorption increases or decreases in different situations. For double quantum wells, it is found that for set of layers more than 8 the peak frequency is inside the terahertz frequency range. For triple quantum wells, it is clearly seen that except for the first case (2 set of layers), the peak frequency within the terahertz range. (For double quantum wells, it is found that by increasing the number of layers more than 20, the peak frequency within terahertz range. In triple quantum wells, it is clearly seen that by increasing the set of layers more than 10, the peak frequency of 10 terahertz or smaller can be obtained.) By adjusting the thickness potential height and the number of the layers, it is possible to adjust the frequency peak of the quantum well structure and the maximum of absorption for application in terahertz sensors.

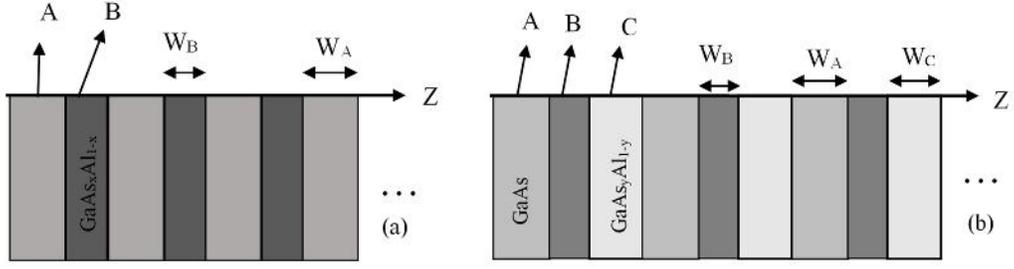

FIG. 1. Schematic diagram for periodic (a) double and (b) triple quantum wells.

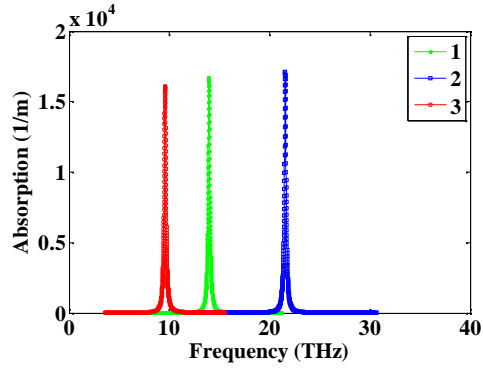

FIG. 2. The absorption spectrum as a function of frequency; curve 1: $N=20$, $W_A=20$ nm, $W_B=20$ nm; curve 2: $N=25$, $W_A=10$ nm, $W_B=15$ nm; curve 3: $N=25$, $W_A=25$ nm, $W_B=25$ nm.

TABLE. 1. The values of parameters of figures 3 and 4

| Figure | $W_A$ | $W_B$ | $N$ | $V_0$ |
|---|---|---|---|---|
| Fig. 3 | $= W_B$ | Variable | Variable | 245.8 meV |
| Fig. 4-a and 4-b | Variable | Variable | 12 | 245.8 meV |
| Fig. 4-c and 4-d | 20 | Variable | 12 | 245.8 meV |

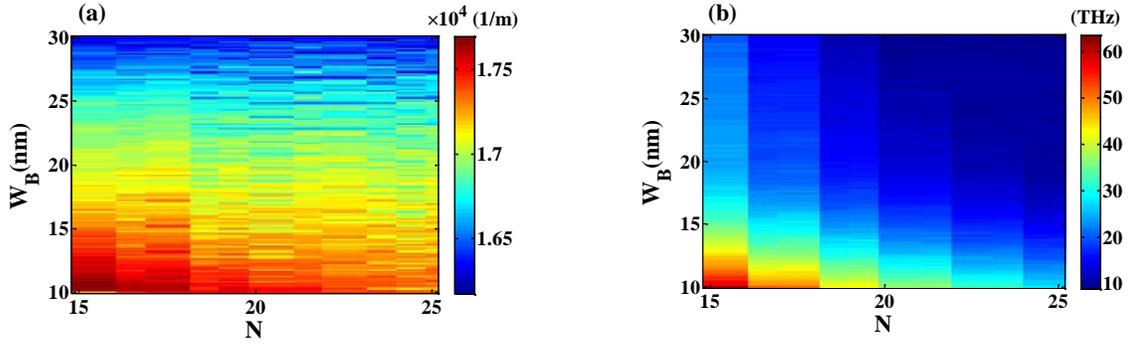

FIG. 3. (a) Maximum absorption coefficient and (b) Frequency of the resonant peak versus the thickness of layer *B* and the number of sets (other parameters are in table 1).

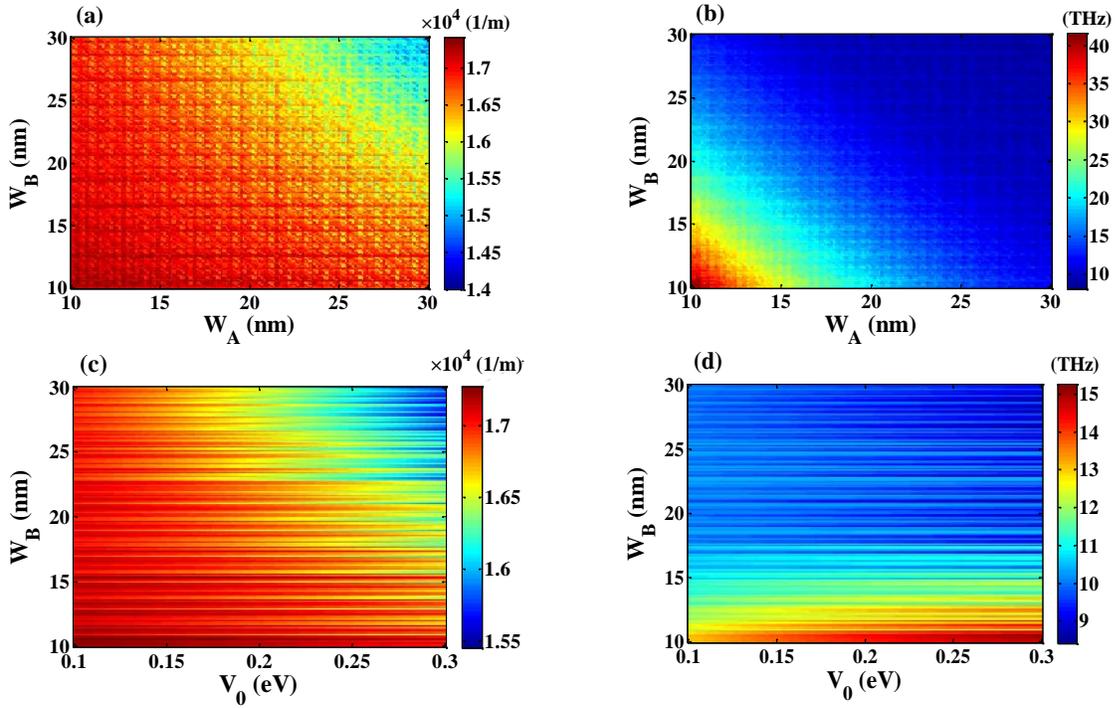

FIG. 4. (a), (b) Maximum absorption coefficient and frequency peak versus the thickness of layers *A* and *B* for 24 layers (c), (d) Maximum absorption coefficient and frequency peak versus the potential height of barrier *B* and the thickness of layer *B* for 24 layers (other parameters are in table 1).

TABLE. 2. The values of parameters of figures 5, 6, 7, 8, 9 and 10

| Figure | $W_A$ | $W_B$ | $W_C$ | $N$ | $V_0$ | $V_0$ |
|---|---|---|---|---|---|---|
| Fig. 5 | 40 nm | 30 nm | 30 nm | Variable | Variable | 0.15 eV |
| Fig. 6 | 40 nm | Variable | 30 nm | Variable | 0.15 eV | 0.15 eV |
| Fig. 7-a and 7-b | 20 nm | 20 nm | 20 nm | 10 | Variable | Variable |
| Fig. 7-c and 7-d | 24 nm | 24 nm | 24 nm | 10 | Variable | Variable |
| Fig. 7-e and 7-f | 30 nm | 30 nm | 30 nm | 10 | Variable | Variable |
| Fig. 8 | 40 nm | Variable | 30 nm | 7 | Variable | 0.15 eV |
| Fig. 9-a and 9-b | 30 nm | Variable | Variable | 8 | 0.15 eV | 0.15 eV |
| Fig. 9-c and 9-d | 30 nm | Variable | Variable | 14 | 0.15 eV | 0.15 eV |
| Fig. 10-a and 10-b | 20 nm | 20 nm | 20 nm | 8 | Variable | Variable |
| Fig. 10-c and 10-d | 20 nm | 20 nm | 20 nm | 14 | Variable | Variable |

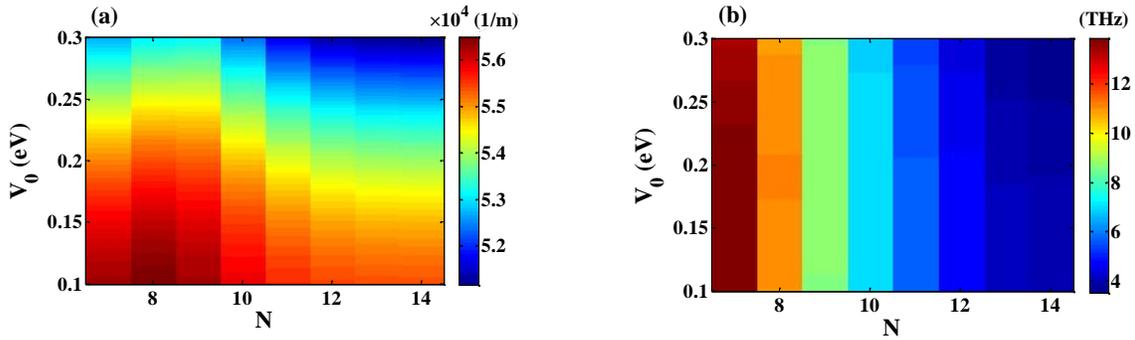

Fig 5: (a)Maximum absorption coefficient and (b)the frequency peak versus the potential height of the barrier *B* and the number of layers, (other parameters are in table 2).

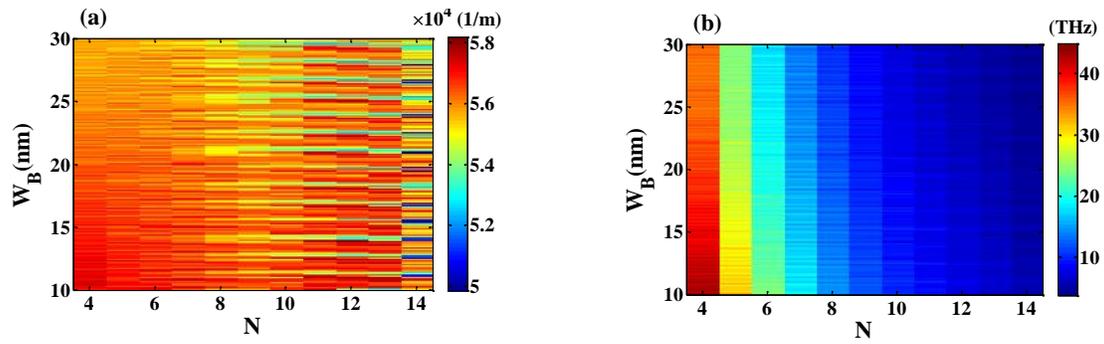

FIG. 6. (a) Maximum absorption coefficient and (b)the frequency peak versus the thickness of layer *B* and the number of layers, (other parameters are in table 2).

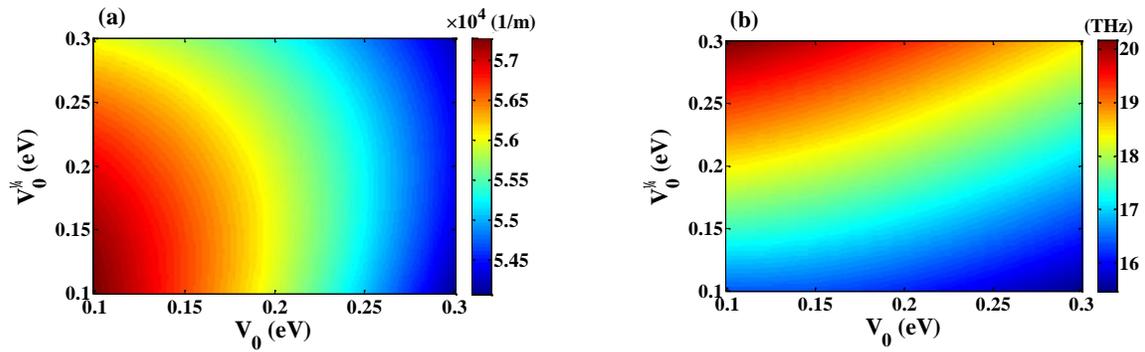

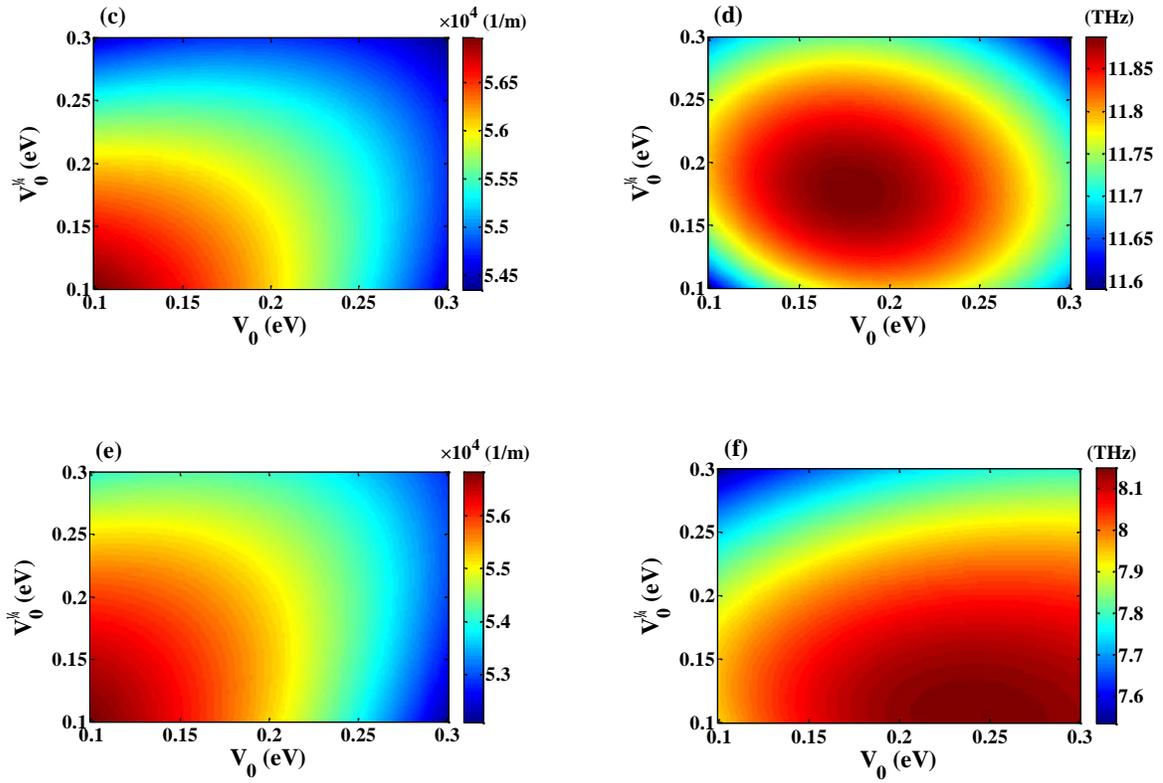

FIG. 7. Maximum absorption coefficient versus the potential height of barrier *B* and *C* (other parameters are in table 2).

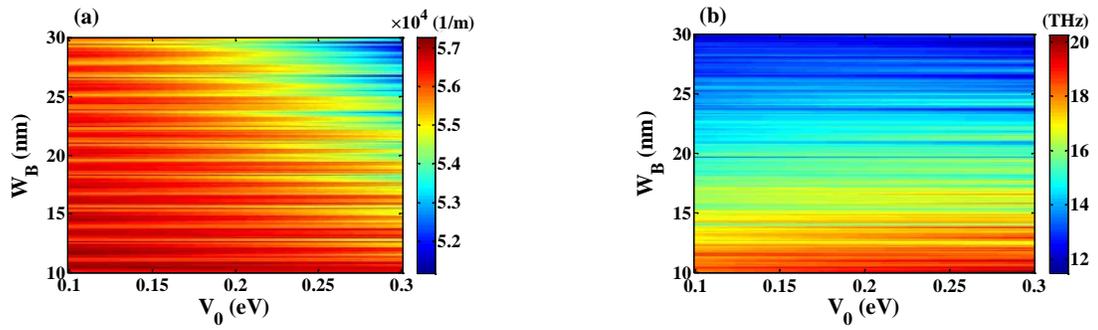

FIG. 8. (a)Maximum absorption coefficient and (b) the frequency peak versus

$V_0$ and $W_B$ (other parameters are in table 2).

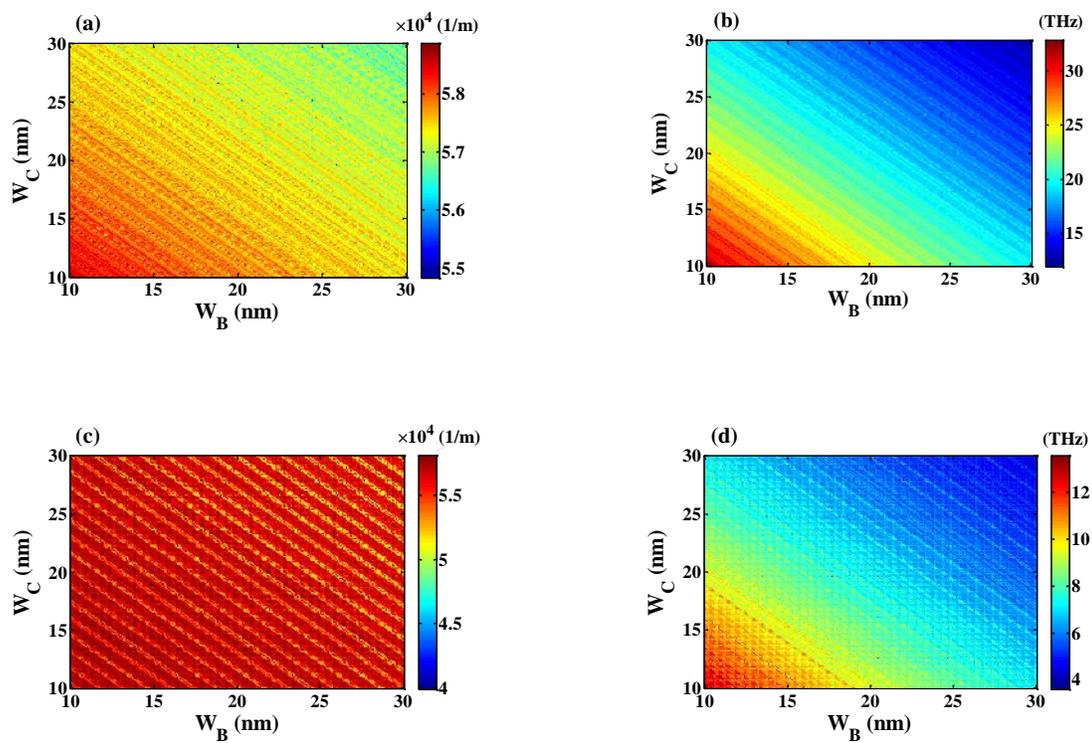

FIG. 9. (a, c) Maximum absorption coefficient and (b, d) frequency peak value versus the thickness of layers *B* and *C* (other parameters are in table 2).

18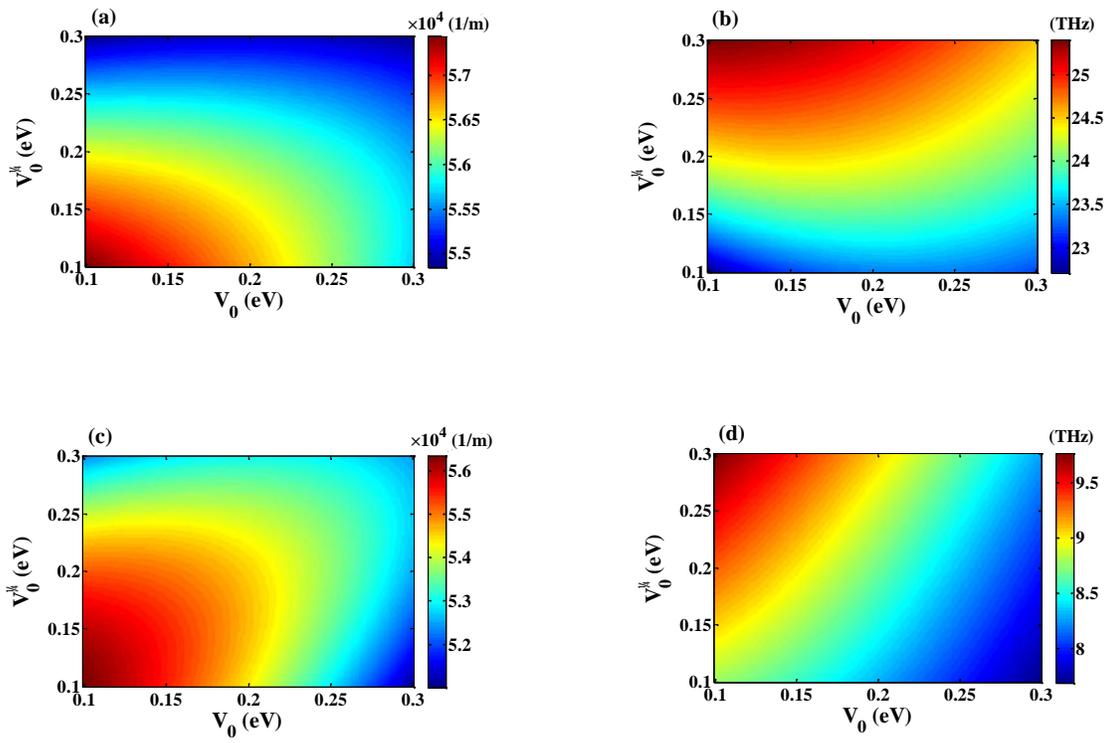

Fig 10. (a, c) Maximum absorption coefficient and (b, d) frequency peak value versus potential height of barriers *B* and *C* (other parameters are in table 2)